# Investigation of measuring hazardous substances in printed circuit boards using the micro-focus X-ray fluorescence screening[*]

FU Ming-lei (付明磊)[1], D.Irzhak[2], R.Fakhrtdinov[2], M. Grigoriev[2], QUAN Bi-sheng (全必胜)[1], LE Zi-Chun (乐孜纯)[1], D. Roshchupkin[2]

1. College of Science, Zhejiang University of Technology, Hangzhou 310023, China

2. Institute of Microelectronics Technology and High-Purity Materials, Russian Academy of Science, Chernogolovka 142432, Russia

**Abstract**: Printed circuit boards (PCBs) are widely used in most electrical and electronic equipments or products. Hazardous substances such as Pb, Hg, Cd, etc, can be present in high concentrations in PCBs and the degradation and release of these substances poses a huge threat to humans and the environment. To investigation the chemical composition of PCBs in domestic market of China, a practical micro-focus X-ray fluorescence (μ-XRF) system is setup to make the elements analysis, especially for detecting hazardous substances. Collimator is adopted to focus the X-ray emitted from X-ray tube. BRUKER X-ray detector with proportional counter is used to detect the emitted fluorescence from the PCB samples. Both single layer PCB samples and double layers PCB samples made of epoxy glass fiber are purchased from the domestic market of China. Besides, a MC55 wireless communication module made by SIEMENS in Germany is used as the reference material. Experimental results from the fluorescence spectrums of the testing points of PCB samples show that, hazardous substances, mainly Pb and Br, are detected from the welding pads and substrates. In addition, statistical data about the average relatively amount of the main substances in testing points are also illustrated. It is verified that μ-XRF screening offers a simple and quick qualitative measurement of hazardous substances in PCBs.

## 1. Introduction

It is well known that printed circuit boards (PCBs) are widely used in most electrical and electronic products and consist of a heterogeneous mix of organic materials, metals, glass fibers, flame retardants, etc. However, some of these organic substances are toxic, such as brominated flame retardants (BFR), polyvinyl chloride (PVC) and heavy metals [1,2]. Especially, hazardous substances, such as Pb, Hg, Cd, $Cr^{6+}$, polybrominated biphenyls (PBB), and polybrominated diphenyl ethers (PBDE), can be present in high concentrations in PCBs and the degradation and release of these substances poses a hazard to humans and the environment [3]. According to the famous RoHS (The restriction of the use of certain hazardous substances in electrical and electronic equipment) directive, the maximum tolerable mass fractions are 0.1% of Pb, Hg, $Cr^{6+}$, PBB and PBDE, and 0.01% of Cd in homogeneous materials, respectively [4]. But, PCBs are usually compound products with complex structures. It consists of a non-conductive substrate or laminate with the printed circuit conductors upon (or within) the substrate and the components mounted to it (chips, connectors, capacitors, processors, etc) [5]. Hence, it has significant meaning to measure concentrations of elements in PCBs at a simple and quick manner in order to detect the hazardous substances immediately at both

[*] Supported by the International Joint Research Program of China (No. 2012DFR10510) and the Research Fund for the Doctoral Program of Higher Education of China (No. 20133317110006)

1) lzc@zjut.edu.cn

2) rochtch@iptm.ru

producing and recycling process.

Recently, elemental analysis of PCBs, especially PCBs from mobile phones and personal computers [4,5,6], has attracted much attention. X-ray fluorescence spectrometry (XRF) including micro-focus X-ray fluorescence (μ-XRF) [3,4,6,7,8,9], instrumental neutron activation analysis (INAA) [3], inductively coupled plasma - optical emission spectrometry (ICP-OES) [3,4,6] and cold vapour - atomic fluorescence spectrometry (CV-AFS) [4] are commonly adopted to make qualitative and quantitative determination of the concentrations of hazardous substances in PCBs. Among them, XRF screening offers a simple qualitative and, to a limited extent, quantitative measurement of hazardous substances in PCBs. An average measurement by XRF takes only a few minutes compared to other methods, such as ICP-OES, which can take up to several hours to get the measurement data due to time-consuming sample preparation [8-11]. Therefore, we choose the XRF screening as the analytical method for measuring hazardous substances in PCBs.

For investigation of the chemical composition of PCBs in domestic market of China, we build a practical μ-XRF system to make the elements anaylsis. We use collimator to focus the X-ray emitted from X-ray tube and BRUKER X-ray detector with proportional counter to record the emitted fluorescence from the PCB samples. In section 2, structure of the μ-XRF system is demonstrated. In section 3, the simple procedures for preparation the PCB samples are introduced. In section 4, experimental results from the fluorescence spectrum analysis are discussed. In section 5, the paper is concluded.

**2. Setup of the μ-XRF system**

Energy dispersive XRF system and μ-XRF system are commonly adopted to make the elemental analysis of PCBs. Difference between the two systems is that X-ray focusing optics should be used in the latter one, where X-ray polycapillary optics is the most popular choice [12-17]. The advantage of μ-XRF system is achieving better limit of detection (LoD). LoD in XRF characterizes the lowest concentration of an element in the sample which is distinguishable as a fluorescence peak over the spectral background noise [11]. According to the standard the limit of detection, $c_{LoD}$, is defined as follows:

$$c_{LoD} = \frac{I_{LoD} - \overline{I_{bl}}}{m_{cal}} \quad (1)$$

Where $I_{LoD}$ denotes the measured signal at the limit of detection for the respective element, $\overline{I_{bl}}$ denotes the mean value of the measured signal of the blank specimen and $m_{cal}$ denotes the slope of the calibration curve [11].

In this paper, we setup a practical μ-XRF system, as shown in Figure 1. Both the X-ray tube and the X-ray focusing optics can be changed in order to optimize the excitation conditions and improve the performances of the method for particular cases. The type of X-ray detector is proportional counter, which is suitable for the PCB measurement. Other types of detectors can also be adopted. For example, Dill *et al* [18], compare the performance of XRF instruments with different detector systems (proportional counter, positive intrinsic negative and Si drift detectors) for measuring thin Au and Pd coatings on PCBs and to investigate different ways of background treatment.

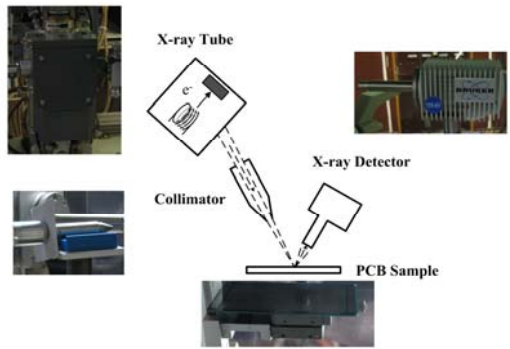

**Figure 1**: Demonstration of the XRF system adopted in this paper. The tube voltage is 50kV and tube current is 150mA. Collimator is used to focus the X-ray beam to achieve a micro focal spot. A BRUKER X-ray detector is used to detect the emitted fluorescence from the PCB samples.

## 3. Sample preparation for XRF screening

XRF is famous for its non-destructive screening, and there is almost no sample preparation required for measurement. But, properly sample preparation can improve the accuracy of measurement. For example, Gore et al [3], analyze different preparation methods (various forms of shredding and milling on samples) of electro-technical products for reduction of hazardous substances compliance testing and conducted a comparative study using four energy dispersive XRF instruments (micro-spot, hand-held, bench-top, and laboratory polarizing XRF). Their work show that fine shredding is probably an optimal form of sample preparation, which suited to the accuracy required for compliance testing. Hirokawa et al [7], proposes a XRF method for the quantitative analysis of Co, Ni, Pd, Ag, and Au in PCBs by using the loose-powder method and correction by scattered X-ray intensity. Their work improves sample preparation for an accurate direct determination.

In this paper, we choose two PCBs from domestic market of China as the measuring targets, as shown in Figure.2. These types of PCBs are widely used in electrical and electronic equipments or products. Considering that there is no components (chips, connectors, capacitors, processors, etc) mounted to PCBs, we choose to measure the welding pads and substrate of the two PCBs. Hence, the PCBs are cut into pieces of dimensions approximately 1 cm plus 1 cm after the testing points are determined.

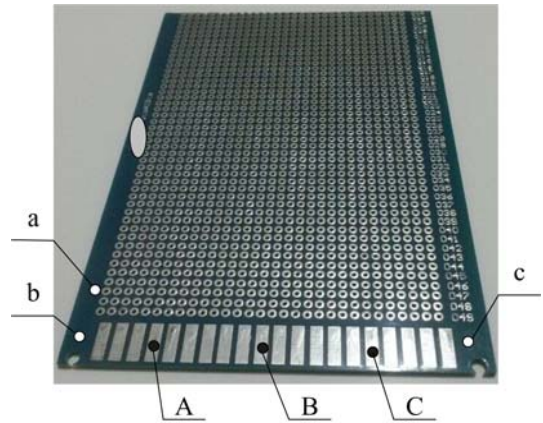

(a) A single layer PCB sample

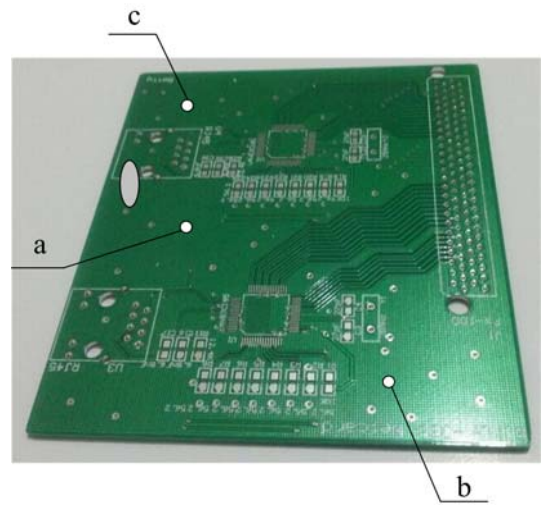

(b) A double layers PCB sample

**Figure 2**: The PCB substrate is made of epoxy glass fiber (FR-4). Three testing points labeled as point A, B and C are used to be analyzed the substances at the welding pads, while point *a*, *b* and *c* are used to be analyzed the substances in the substrate. Besides, for the double layers PCB samples, there are coatings of copper clad laminates at top and bottom layer.

We also choose the MC55 wireless communication module as the reference material, as shown in Figure.3. The module conforms to the RoHS directive.

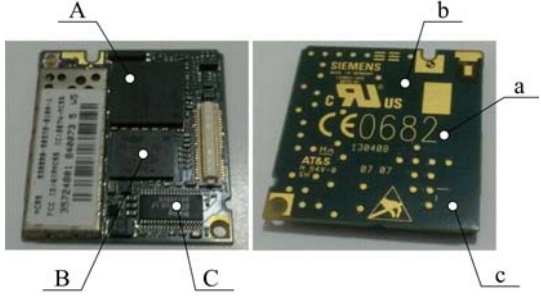

**Figure 3**: The MC55 wireless communication module made by SIEMENS in Germany. Three testing points labeled as point A, B and C are used to be analyzed the substances in the chips, while point *a*, *b* and *c* are used to be analyzed the substances in the substrate.

## 4. Analysis of the X-ray fluorescence spectrum

By the XRF spectrum, a linear relationship between the measured fluorescence intensity $I_i$ of the line $i$ and the mass fraction $c_i$ of the corresponding element in the sample can be assumed, if absorption and enhancement effects are neglected [19].

$$I_i = const_i \cdot c_i \tag{2}$$

The calibration constant $const_i$ consists of various parameters like the excitation intensity and the probability for the production of the fluorescence line $i$ under the given experimental conditions. Hence, we calculate the relative quantity of the substances in PCBs according to the counts number from detector for simplicity.

$$I_i^R = \frac{count_i}{\sum_{i=1}^{n} count_i} \tag{3}$$

In formula (3), $count_i$ denotes the counts number of certain substance. $n$ denotes the total number of substances that have been detected.

Firstly, we make XRF analysis on the welding pads of PCBs. Figure 4 shows the fluorescence spectrum of testing point A in Figure 2(a). Cu and Sn are commonly used in welding pads. But, Pb and Br which are hazardous substances restricted by RoHS are also detected. Au does not come from welding pads. It is emitted from anode material of X-ray tube. For the same reason, Au $L_\alpha$ line and $L_\beta$ line can be found in all the fluorescence spectrums in this paper.

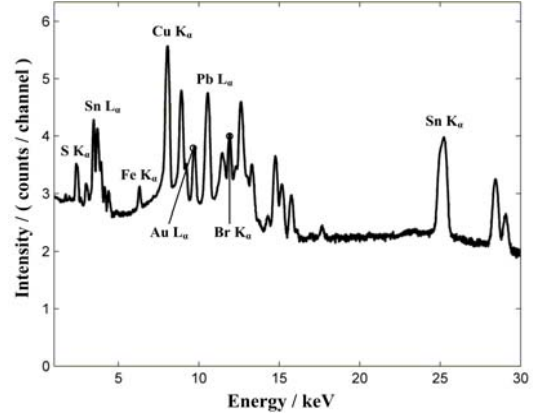

**Figure 4**: The fluorescence spectrum of testing point A in Figure 2(a). The logarithmic coordinates are adopted by the vertical axis. Apparently, hazardous substances Pb and Br are detected.

Although XRF can only offer a simple qualitative or semi-quantitative analysis for the substances of the samples, it is still meaningful to make the chemical composition of the testing points about the relatively amount of the main substances. As shown in Figure 5, Pb and Br are 12.05% and 2.33% of the substances in welding pads of PCB sample. Note that XRF screening can only measure total Br and cannot discriminate between polybrommated diphenyl ether (PBDE) and polybrominated biphenyl (PBB). The same applies to Cr, and it cannot distinguish between hexavalent and trivalent chromium [8].

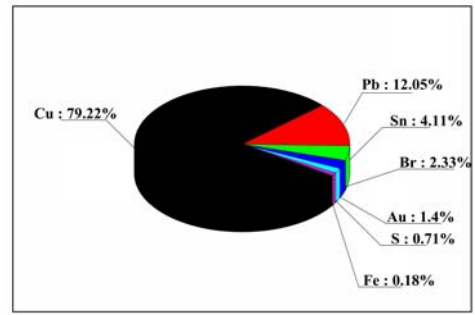

**Figure 5**: Average relatively amount of the main

substances in testing points

Secondly, we make XRF analysis on the substrates of PCBs. Figure 6 shows the fluorescence spectrum of testing point *a* in Figure 2(a). Obviously, Br has a high concentration and Pb is also detected.

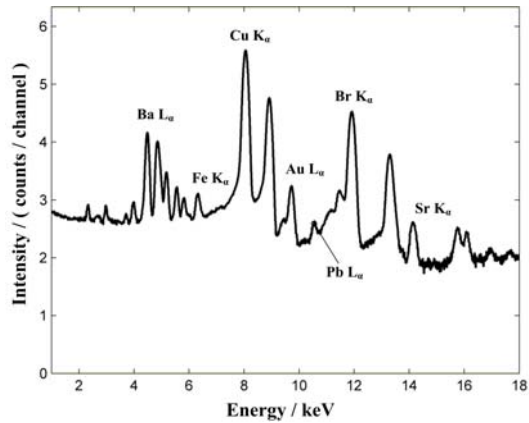

**Figure 6**: The fluorescence spectrum of testing point *a* in Figure 2(a). The logarithmic coordinates are adopted by the vertical axis. Apparently, hazardous substances Br and Pb are detected.

As shown in Figure 7, Br and Pb are 7.68% and 0.1% of the substances in substrates of PCB sample.

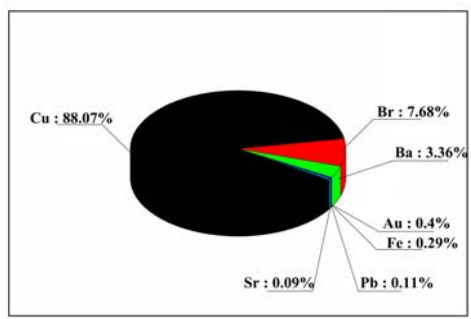

**Figure 7**: Average relatively amount of the main substances in testing points

Thirdly, we make XRF analysis on the substrates of the PCBs sample in Figure 2(b). The hazardous substance Br has the highest concentration of all the substances.

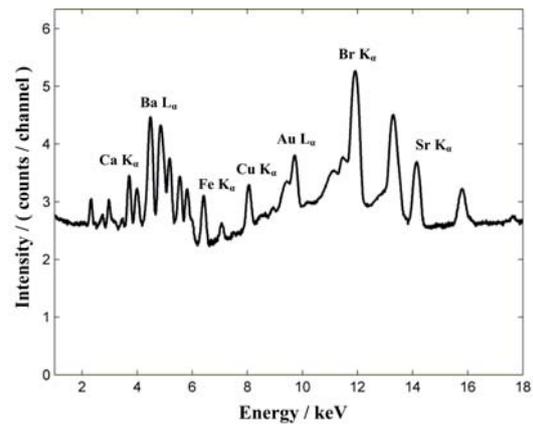

**Figure 8**: The fluorescence spectrum of testing point *a* in Figure 2(b). The logarithmic coordinates are adopted by the vertical axis. Apparently, hazardous substance Br is detected.

As shown in Figure 9, 79.87% of the substances in substrate of PCB sample is Br.

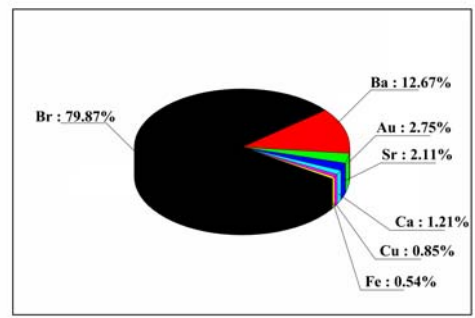

**Figure 9**: Average relatively amount of the main substances in testing points

To make the comparison between the PCB samples with the formal qualified product, we also make XRF analysis on the MC55 module under the same conditions. The fluorescence spectrum of the testing chip is shown in Figure 10. No hazardous substance is detected.

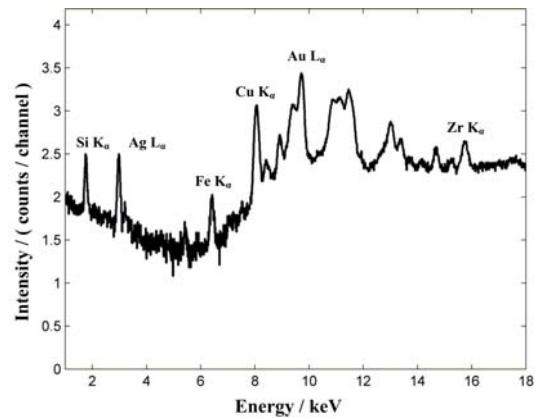

**Figure 10**: The fluorescence spectrum of testing point A in Figure 3. The logarithmic coordinates are adopted by the vertical axis. No hazardous substance is detected.

The fluorescence spectrum of the testing substrate is shown in Figure 11. Also, no hazardous substance is detected.

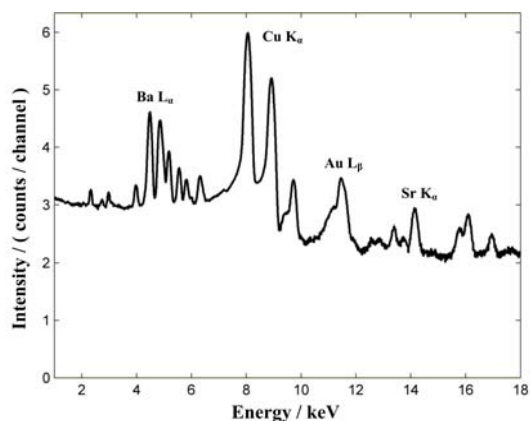

**Figure 11**: The fluorescence spectrum of testing point *a* in Figure 3. The logarithmic coordinates are adopted by the vertical axis. No hazardous substance is detected.

Table 1 shows the summary of XRF analysis results for the samples. All the testing points of the samples are measured at the same condition. In addition, to obtain stable fluorescence spectrum, the measuring time for each points is 30 minutes. According to Table 1, both Pb and Br are detected in the PCB samples. In the substrate of the double layer PCB sample, Br has extremely high concentration. Considering that the PCBs which are the same or similarly with the samples are widely used in domestic market of China, it is quite a serious problem need to be concerned.

**Table 1**: Results of the XRF analysis for the samples

| Samples | Pb | Br | Hg | Cd | $Cr^{6+}$ |
|---|---|---|---|---|---|
| Welding pads of single layer PCB | 10.22% | 3.37% | - | - | - |
| Substrate of single layer PCB | 0.12% | 6.52% | - | - | - |
| Substrate of double layers PCB | - | 63.54% | - | - | - |
| Chips of MC55 | - | - | - | - | - |
| Substrate of MC55 | - | - | - | - | - |

## 5. Conclusion

The electronics industry has become one of major growth points in economy in China recently, especially in coastal areas. And PCBs are widely used in most electrical and electronic equipments or products. It is very meaningful to monitor the quality of PCBs in order to reduce the release of hazardous substances, especially those PCBs mainly used in the domestic markets. However, PCBs are usually compound products with complex structures. It probably costs too much time for the element analysis methods which need fine sample preparations. In this paper, we verify that μ-XRF screening might be the one of the best methods to make the element analysis for PCBs. Experimental results also show the typical hazardous substances Pb and Br are detected in the PCB samples. Another potential advantage for μ-XRF screening is that both the X-ray tube and the optics can be changed quite easily in order to optimize the excitation conditions and improve the performances of the method for particular cases. Still, limit of detection (LoD) of μ-XRF screening for PCBs is the main problem for further studies.